\documentclass[preprints,article,accept,moreauthors,pdftex]{mdpi}

\firstpage{1} 
\pubvolume{xx}
\issuenum{1}
\articlenumber{5}
\pubyear{2020}
\copyrightyear{2020}
\history{}

\usepackage{listings}
\usepackage{subfiles}
\usepackage{subfigure}
\usepackage{amsmath}
\usepackage{url}
\def\UrlBreaks{\do\/\do-}
\usepackage{breakurl}
\usepackage{forest}
\forestset{qtree/.style={for tree={parent anchor=south, 
           child anchor=north,align=center,inner sep=2pt}}}
\usepackage{multirow}

\Title{A survey on applications of augmented, mixed and virtual reality for nature and environment}

\Author{Jason Rambach $^{1,\dagger}$*, Gergana Lilligreen $^{2,\dagger}$, Alexander Schäfer $^{3,\dagger}$, Ramya Bankanal $^{3}$, Alexander Wiebel $^{2,}$ and Didier Stricker$^{1,3}$}

\AuthorNames{Jason Rambach, Gergana Lilligreen, Alexander Schäfer, Ramya Bankanal, Alexander Wiebel, Didier Stricker}

\address{%
$^{1}$ \quad German Research Center for Artificial Intelligence, Augmented Vision department, Kaiserslautern, Germany; Jason.Rambach@dfki.de\\
$^{2}$ \quad Hochschule Worms, UX-Vis Group, Worms, Germany ;  lilligreen@HS-Worms.DE\\
$^{3}$ \quad University of Kaiserslautern, Kaiserslautern, Germany}

\corres{Correspondence: Jason.Rambach@dfki.de}
\firstnote{These authors contributed equally to this work.} 

\abstract{Augmented reality (AR), virtual reality (VR) and mixed reality (MR) are technologies of great potential due to the engaging and enriching experiences they are capable of providing. Their use is rapidly increasing in diverse fields such as medicine, manufacturing or entertainment. However, the possibilities that AR, VR and MR offer in the area of environmental applications are not yet widely explored. In this paper we present the outcome of a survey meant to discover and classify existing AR/VR/MR applications that can benefit the environment or increase awareness on environmental issues. We performed an exhaustive search over several online publication access platforms and past proceedings of major conferences in the fields of AR/VR/MR. Identified relevant papers were filtered based on novelty, technical soundness, impact and topic relevance, and classified into different categories. Referring to the selected papers, we discuss how the applications of each category are contributing to environmental protection, preservation and sensitization purposes. We further analyse these approaches as well as possible future directions in the scope of existing and upcoming AR/VR/MR enabling technologies.}


\begin{document}


\section{Introduction}
A healthy environment is vital for all living beings. Protecting and preserving our environment is necessary as natural resources are limited. The need for a change in paradigm has become apparent in the last years~\cite{ipcc_1.5}. It is clear now that apart from the support of governments and industry, it is each individual’s responsibility to make correct use of resources, conserve energy and adjust their lifestyle to support sustainability.\par
In the last couple of decades we have witnessed a tremendous number of new technologies that have made life easier and better. Augmented reality (AR), virtual reality (VR) and mixed reality (MR) are technologies that either create fully virtual worlds (VR) or combine virtual elements with the real world (AR, MR). The enhancement of the real world environment by superimposing virtual content has the potential of providing enriching, interactive and captivating experiences. Highly interesting applications at a level of proof of concept systems or even commercial applications have been introduced in many diverse fields such as medicine and health care \cite{chen2017recent,qian2015surgery}, education \cite{jensen2018review,bower2014augmented,highereducation2019}, entertainment \cite{von2017augmented,liszio2016vrgaming}, industrial maintenance \cite{DoFarticle} and many more.
In this survey we investigate the use of AR, VR and MR in applications that relate to the preservation of the environment in a \emph{direct} or \emph{indirect} manner. It is apparent that such ecological applications are currently few in number, making this a topic that has not been explored extensively yet. Within this work we attempt to discover existing publications on the topic and classify them in a meaningful way. We look into applications that have a \emph{direct} connection to ecology, but also into AR/VR/MR technologies that have the potential of contributing positively to the state of the environment in an \emph {indirect} way, such as e.g. through the reduction of transportation needs for business purposes.\par
To the best of our knowledge, this is the first survey to cover the particular topic of AR/VR/MR for ecology and nature. Apart from listing existing work in the field, our survey investigates the topic and uncovers the potential opportunities that come with it. We look at the current state of AR/VR/MR in terms of existing technologies for tracking, display and scene understanding and make an initial evaluation of how upcoming advancements in these areas can contribute to the success of ecological AR/VR/MR topics.
The rest of the paper is organized as follows: We start by describing the exact methodology (i.e. keywords, sources, selection criteria) we used for finding existing work relevant to this survey in Section \ref{section:Methodology}. The classification structure that we propose is presented in Section \ref{section:Classification} and the respective categories are explained. Statistics on the selected publications are provided in Section \ref{section:Stats}. In Section \ref{section:Work}, we describe the main points of the most prominent work from each of the categories. In Section \ref{section:Ideas}, we summarize our findings and also discuss the current state of AR/VR/MR and possible future developments.
Concluding remarks are given in Section \ref{section:Conclusion}.

\section{Search Methodology}
\label{section:Methodology}

\begin{figure}[h]
\centering
\includegraphics[width=0.6\columnwidth]{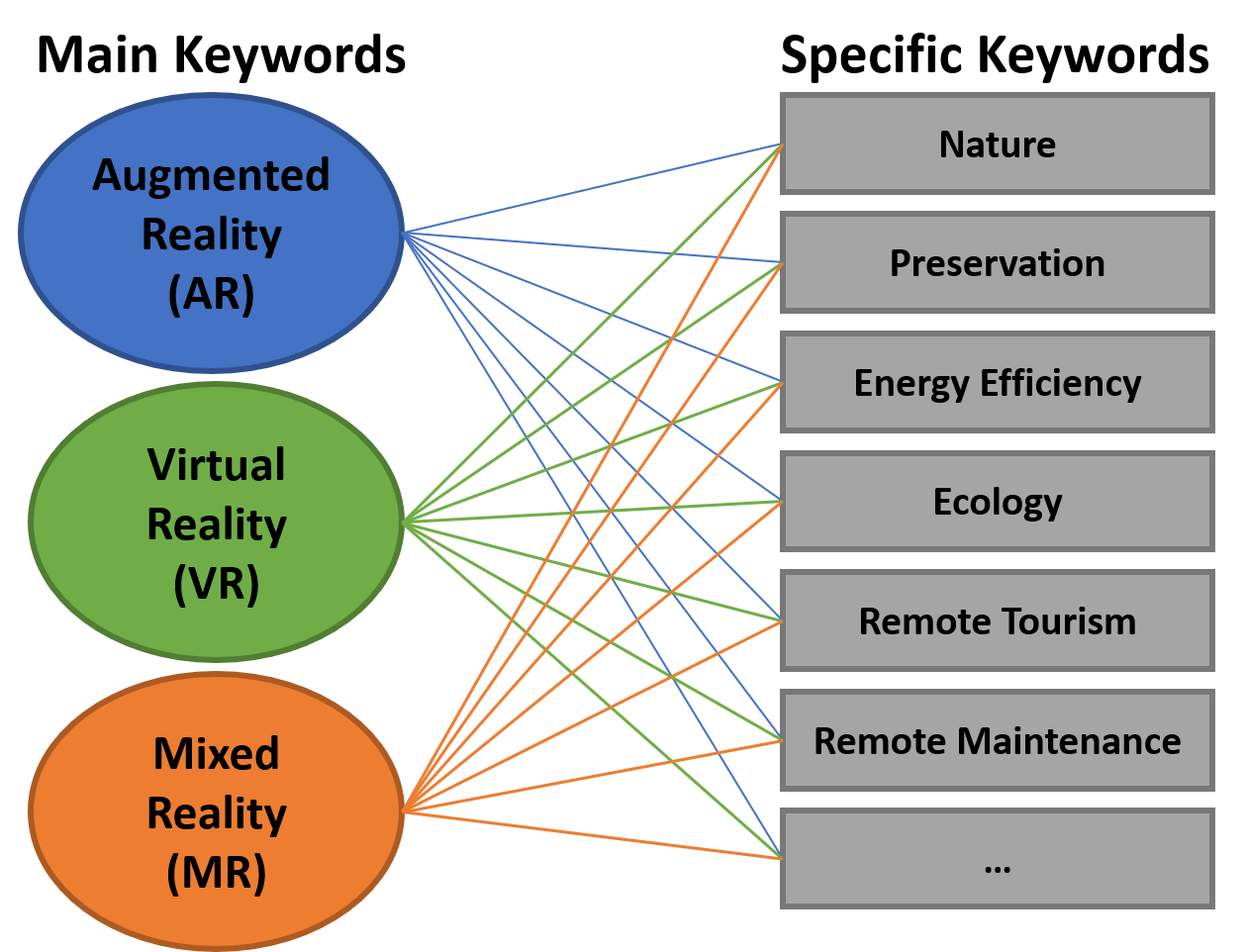}
\caption{Examples of keyword combinations used in our search.}
\label{keywords}
\end{figure}

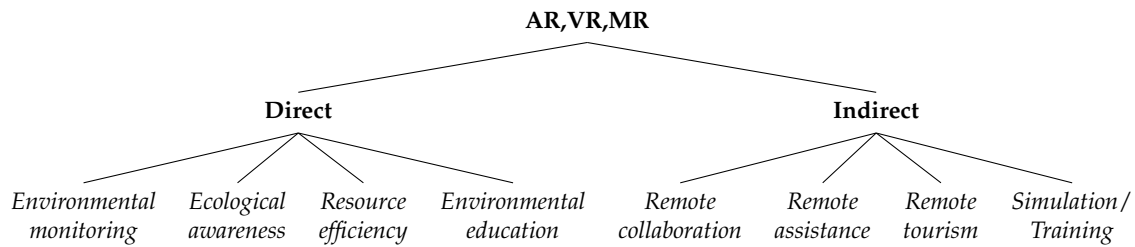
\begin{figure*}[htb]
    \large
    \centering
    \begin{forest}, baseline, qtree
[{\textbf{AR,VR,MR}} 
[\textbf{Direct},
    [\textit{Environmental}\\\textit{monitoring}]
    [\textit{Ecological}\\\textit{awareness}]
    [\textit{Resource}\\\textit{efficiency}]
    [\textit{Environmental}\\\textit{education}]]
[\textbf{Indirect},
    [\textit{Remote}\\\textit{collaboration}]
    [\textit{Remote}\\\textit{assistance}]
    [\textit{Remote}\\\textit{tourism}]
    [\textit{Simulation}/\\\textit{Training}]
]]
 \end{forest}
    \caption{Classification of augmented, virtual and mixed reality environmental applications.}
    \label{fig:classification_label}
\end{figure*}
We adopted an extensive search approach to discover as many existing publications as possible on the topics of interest. The search was performed using search queries in different data sources.  Prominent data sources were Scopus (\url{https://www.scopus.com}), Google Scholar (\url{https://scholar.google.com/}) and IEEE Xplore (\url{https://ieeexplore.ieee.org/}). Apart from searching these databases we also explored recent proceedings of leading AR/VR/MR conferences such as ISMAR, IEEE VR, EuroVR. 
Since  AR/VR/MR are used in numerous and diverse fields, the challenge was to define a set of suitable keywords from different domains relevant to the topic. The search was performed by combining keywords that relate to AR/VR/MR ("augmented reality" OR "mixed reality" OR "virtual reality" OR "AR" OR "MR" OR "VR") with keywords relating to specific environmental topics to form regular expressions. A list of several of the keywords used is shown in Figure \ref{keywords}. All possible combinations of keywords from the left column with keywords from the right column were tested. An initially short list of keywords was expanded by the findings of the search. Through this iterative procedure, the search finally led to the initial identification of over 300 related publications. The literature search was performed between November 2019 and February 2020.\par 
We decided to address the topic from both a more narrow and a broader perspective. Therefore the results of the search contained publications that explicitly focus and mention environmental issues (\emph{direct} applications class; narrow) and publications that present applications that can have an environmental impact (based on the opinion of the surveyors) which is however not explicitly stated by the authors of the publication (\emph{indirect} applications class; broad). 
\par
It became apparent at early stages of the search that topics covered in the indirect class have such a large extent of related publications that each topic would require a dedicated survey. This exceeds the scope of this survey. Thus, it has been decided to exhaustively search the direct area of environmental applications (narrow) only. For the indirect applications area, we present some key publications which from our point of view support the discussion on environmental perspectives. Therefore, the indirect application part of this survey is based on a non-exhaustive (broad) search. 
The next step was the filtering of the search results in order to decide on the most influential papers for further study and to initiate the classification procedure. The main criteria for performing the filtering step were:
\begin{itemize}
    \item \textbf{Relevance:} We graded the papers based on how closely the proposed work is related to the specific topic of environmental AR/VR/MR.
    \item \textbf{Originality:} In case of papers addressing similar topics with similar methodology, we kept the earlier paper and discarded later ones, unless they present an important advancement to the state-of-the-art.
    \item \textbf{Impact:} When we found too many papers on the same topic, we took the number of times that a paper was cited by subsequent publications as a measure of its impact. Additionally, the quality of the conference or journal where the work appeared was taken into consideration.
\end{itemize}
The publications were graded individually by the three main surveyors (score 0-10) in each category based on the paper title and abstract. The final selection was done after a discussion round between all 3 main surveyors. A threshold of an average score of $5.0$ was set for selection. Additionally, a score below $2.5$ in one of the categories led to direct exclusion. For the direct area of environmental AR/VR/MR publications this led to selecting 28 papers. Very few papers were excluded in the direct category, which shows that the amount of work in the field is still extremely limited. Each selected paper was read fully by at least one of the main surveyors. 
The selected papers were classified into several categories as discussed in the next section. Papers are discussed in more detail within the description of each classification category.

\section{Classification}
\label{section:Classification}

Our proposal for classification of environmental AR/VR/MR applications is presented in Figure~\ref{fig:classification_label}. Publications were first classified into the categories \emph{direct} and \emph{indirect} based on their type of relation to environmental topics. The category \emph{direct} encompasses all applications and methodologies that are associated in a direct way with nature and ecology. On the other hand, the category \emph{indirect} contains applications and methodologies with a looser connection to environmental issues. Applications from different domains that can have a clear impact on environmental issues like reduction of travel and its subsequent reduction of emissions are covered here. These approaches might not engage with the environment directly but help to reduce wasting of vital resources and pollution.\par
The category \emph{direct} involves subcategories like environmental education and ecological awareness, in which AR/VR/MR applications provide illustrations of characteristics and other related information about species that exist around us. As they learn more about nature and its problems, people can be encouraged to change their behavior and become more ecologically responsible. Environmental monitoring is another major category here. This category covers applications that visualize and monitor data on climate-related topics like air, water and soil. AR as an engaging tool, is capable of illustrating the negative influence of pollution, climate change etc. in order to make their effects fully tangible by humans. Finally, AR/VR/MR applications that support efficient usage of resources are placed under the resource efficiency category.\par 
The category \emph{indirect} involves subcategories like remote collaboration, remote assistance and remote tourism. In all applications from the subcategories there is an underlying potential of saving resources through the reduction of the need to travel for professional or entertainment reasons. AR/VR/MR as highly engaging techniques are able to create immersive remote experiences that help in achieving this. In the next section we look into each one of the categories individually.

\section{Statistical Analysis\label{section:Stats}}
\begin{figure}[h]
\centering
\includegraphics[width=1\textwidth]{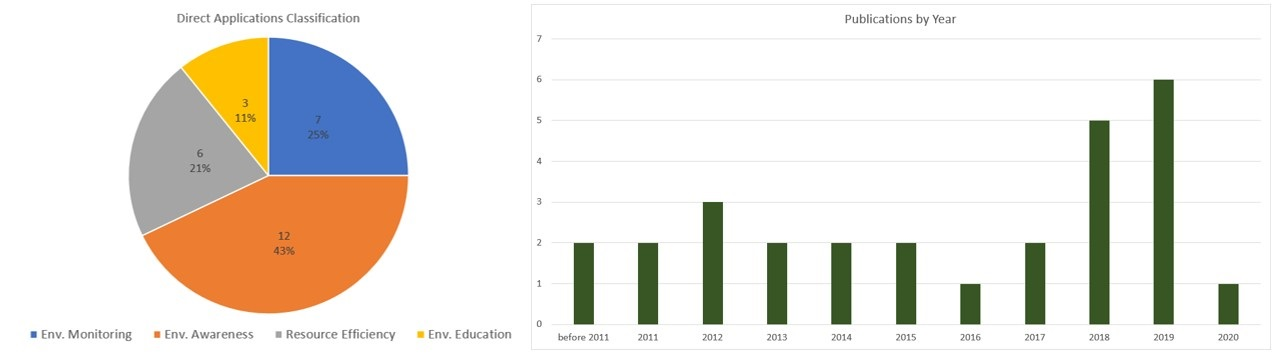}
\caption{Statistical analysis of publications cited in this survey in the \emph{direct} environmental applications category. Left: Distribution by application class. Right: Distribution by year. }
\label{fig:stats}
\end{figure}
A statistical analysis of the publications mentioned in this survey is relevant only for the direct environmental application category which was researched in an exhaustive manner. A total number of 28 papers was selected in this category. In Figure \ref{fig:stats}, we show a pie-chart of the distribution of selected papers over the four classification categories of direct applications on the left side. On the right side, we show the distribution of papers based on the year of publication. All years up to 2010 were summarized in one category. For 2020 the result is only partial as the search for papers ended in February 2020. Of the 28 papers, 12 ($42.9\%$) were published as journal articles and 18 ($57.1\%$) in conference proceedings. It is also worth mentioning that only 4 of the 28 publications were done in dedicated AR/VR/MR journals or conferences.\par
We have also analyzed the number of citations of publications based on information from Google Scholar (\url{https://scholar.google.com/}) in August 2020. The average number of citations over the 28 selected publications is $10.42$. The 5 most cited publications are \cite{Markowitz2018}(73), \cite{envmonitoring2013}(40), \cite{Neugebauer2011}(35), \cite{Deleon98virtualflorida}(26), \cite{romao2002}(23). In citations per year the 5 most cited publications are: \cite{Markowitz2018}($24.33$), \cite{Fan2018}($5.33$), \cite{envmonitoring2013}($5$), \cite{MYLONAS201989}($5$), \cite{Neugebauer2011}($3.5$).
The average number of citations for each classification category is as follows: environmental monitoring $9.86$ (most cited \cite{envmonitoring2013}(40)), ecological awareness $12.25$ (most cited \cite{Markowitz2018}(73)), resource efficiency $10.83$ (most cited \cite{Neugebauer2011}(35)), environmental education $3.66$ (most cited \cite{huang2019ecologicaldata, zoo14}(5)).

\section{Main Survey Insights}
\label{section:Work}
\subsection{Direct Ecological Applications}
This section presents a state-of-the-art analysis of AR/VR/MR applications with a direct connection to environment and nature for tasks such as monitoring, resource management or education.   

\subsubsection{Environmental Monitoring}
Extreme changes in the environment fostered the development and usage of sensor networks for environmental data. The visualization of such data is an important issue. For Veas et al.\cite{envmonitoring2013} environmental monitoring is “the process of continuously observing and regularly measuring environmental parameters of a specific area in order to understand a phenomenon”. The authors present a multilayer infrastructure and a mobile AR platform that leverages visualization of geo-referenced sensor measurements and simulation data in a seamless integrated view of the environment. The main components in the infrastructure are sensor and acquisition components, as well as simulation and analysis components. The mobile support is divided into deployment, providing infrastructure to access data and prepare it for rendering, and a run-time client, providing visualization and interaction capabilities. Two scenarios are chosen as case studies for the system - a snow science scenario in the context of snow avalanches and a hydrology scenario in the context of watershed modeling. AR shows a major value for presenting subsurface structures or different layers of "material" like soil, grass or snow. During workshops with experts, the authors noted a higher preference for 3D visualization in general, especially in unfamiliar environments. As an overall result the authors stated that on-site AR environmental monitoring can be regarded as a promising field.

Trees provide oxygen, conserve water, ameliorate quality of air by reducing amount of $CO_2$ and also preserve soil. Thus, reforesting helps in the fight against global warming and natural disasters while supporting wildlife as well. AR along with other technologies can be used for localization and monitoring of forests. In this context, West et al.~\cite{west2012metatree} present work on visualizing and monitoring the urban forests using mobile AR. The authors’ main aim is to provide an enriching experience to users and encourage the public to involve in tree planting and monitoring initiatives. They mention limitations of existing mechanisms such as remote sensing and field-based imaging and inventory and brief on how AR tries to overcome them. The prototype they present allows users to calibrate the tree location then displays the dashboard presenting the diameter at breast height (DBH), height and other parameters of tree. In follow-up research~\cite{west2013rephoto} they investigate the integration of ground level rephotography with available LiDAR data, for creating a dynamic view of the urban forest, and its changes across various spatial and temporal scales. They explore the potential of overlaying augmentations within the picture taking process in a smartphone app for guiding persons to guarantee consistency in measurements over growing seasons.\par
In the PAN project \cite{albers2017} AR is used for environmental monitoring with smartphones in a different setting. Users located near points of interest (POI) can take photographs with an AR application and act as citizen scientists. With an AR approach, they are navigated and guided to take a photo on a specific point and in a specific direction. The photographs are processed to time-lapse videos and visualize the environmental changes and can be used for long term documentation. In their project the AR application is used in a limestone pit application case and for observing and documenting the progress of regularly flooded wetlands for the development of a valuable secondary biotope with many rare animal and plant species.\par
Fan et al. \cite{Fan2018} provide a solution to accurately estimate the tree position, diameter at breast height (DBH) and height using a mobile phone with RGB-D-based Simultaneous Localisation and Mapping (SLAM). The measurements obtained are augmented on-screen using AR. This allows the users to easily observe whether estimated measurements are close to actual ones. The authors conclude that estimation of tree parameters with the proposed method gives accurate results.\par
A mobile AR application (PeakLens) for mountain exploration is discussed by Frajberg \cite{mountainAR17}. The application can be used to crowd-source mountain images for environmental purposes, such as the analysis of snow coverage for water availability prediction and the monitoring of plant diseases. Meta-data about mountain peaks is overlaid on the phone screen in real-time over relevant objects. For the AR module a computer vision algorithm, based on Deep Learning, is used for the extraction of the natural mountain skyline from the camera view of the mobile phone: “…such skyline is aligned with the skyline extracted from a virtual panorama, computed from the GPS position of the user, the orientation of the mobile phone view, and a Digital Elevation Model (DEM) of the Earth”\cite{mountainAR17}. The application can also handle occlusions, so that peaks covered by an object in front of them (e.g., a bell tower, a person, or a tree) are not displayed.\par
A different viewpoint in the category environmental monitoring and AR is described by Studer et al. \cite{water11}. The art and science project fleabag uses data values from a biomonitoring system for river water quality. The data is analyzed and then transformed artistically, using AR, to expand the human perception through a sensually clear, visual and physical presence on the river side. For the visualization a symbolically shaped big 'water drop' is sculpted, offering an insight into the world of water fleas. On the back side of the transparent sculpture, a semi-transparent rear projection screen is embedded to project the animation. With good water quality, the 'good water particles' are floating and forming new patterns. The 'water fleas' (green colored) move in a 'relaxed' way through the pattern. With worse water quality the water fleas are colored in red and try to escape from the “bad” water particles, some of them disappear. Since art is used to attract the attention of people to water pollution level, this installation also raises ecological awareness, which is discussed in the following section. 
\subsubsection{Ecological Awareness}
In recent times we have seen quickly rising pollution levels in many areas. The World Health Organisation states that "Between year 2030 and 2050, climate change is expected to cause approximately 250,000 additional deaths per year, from malnutrition, malaria, diarrhoea and heat stress" \cite{WHOFacts}. Emission of green house gases and deforestation have, among other factors, resulted in increasing global warming. Thus it is essential to provide climate change information and its adverse effects to people in an understandable manner. Visualization of climate data with AR can be an effective approach due to AR's captivating nature. Such visualizations could have the ability to change the perception of the situation and encourage a change in behavior.\par
The work of Ramachandran et al. \cite{Ramachandran2019} and Torres et al. \cite{torres2019} aims to contribute in this regard by developing applications with AR that visualize pollutants present in the air. Santos et al. have developed the application eVision \cite{romao2012behavior}. In addition to gathering information about air quality, the application detects environmental threats like cars or factories. The users can "wipe" over these objects and the app than automatically displays objects that are environmental friendly, e.g. overlaying a car with a bicycle.\par
Environmental simulations allow a safe and easy way to acquire knowledge about hard-to-reach habitats. Taulien et al.\cite{maritime19} presented a MR simulation that enables users to convert a place like their living room into a maritime habitat, using the example of the Baltic Sea. The simulation system uses a HMD (Microsoft HoloLens) and the users can explore a virtual underwater world by walking in the real world. Animals, plants and stones are added and respond to users’ actions. Also, rearranging real world objects causes the virtual world to change its look. Through voice-over, additional information is provided when the users look at an object for a longer time, e.g. the oxygen production of a plant. A knowledge questionnaire for two groups - users using text vs. mixed-reality application - showed that both groups improved their knowledge, with a larger improvement using MR. Interestingly, some participants said that they were so fascinated by the MR simulation that they forgot to listen to the voice-over. Additional results show that the users of the mixed-reality application were more excited to experience the habitat.\par
To raise the awareness on energy consumption, a treasure hunt game where AR is used to visualize effects of energy waste and saving, is presented by Buehling et al. in \cite{energysaving12}. The game content, including problem, solution, and clues, is presented through an AR phone application using markers. For example, AR is used to visualize energetic effects like the loss of heat when cooking with an open pot. Red arrows are overlaid on the open pot showing how energy is lost. In order to increase the motivation, the players’ performance is visually presented as dynamic virtual gardens that change from poor to good status depending on the points collected in the quiz game. The use of markers generates serious restrictions for such as an application.\par
An AR underground visualization application that uses a GPS system and image recognition techniques to localize the user are explained in \cite{romao2002}. Geo-referenced information is used to augment virtual underground objects and monitor the water quality or the subsoil structure to locate infrastructures for public supply networks like water, sewage and telephone.\par
\begin{figure}[h]
\centering
\includegraphics[width=0.9\columnwidth]{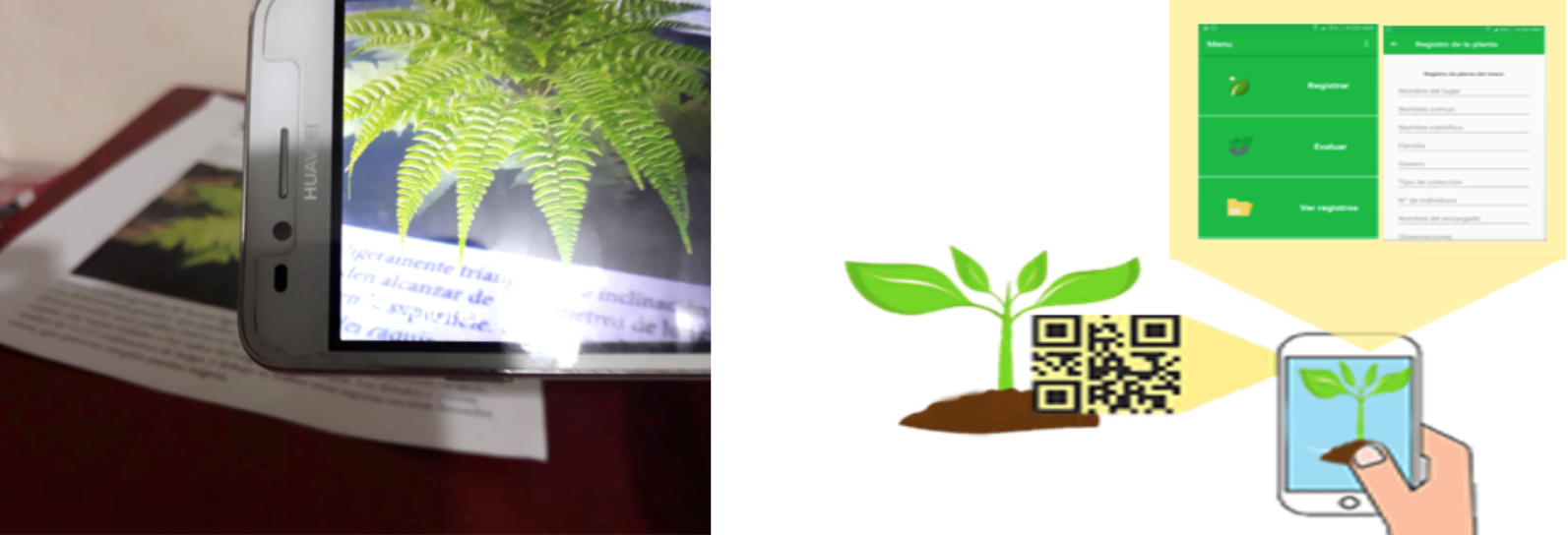}
\caption{Fern visualized using AR on a mobile device (left) and list of its plant characteristics (right). Picture from \cite{Rico_Bautista_2019} available via Creative Commons Attribution 3.0 Unported license (\url{https://creativecommons.org/licenses/by/3.0/}).}
\label{fig:digitalalbum}
\end{figure}

Botanical gardens preserve, protect and display a wide range of flora species. An interactive learning AR mobile application was developed for a botanical garden in \cite{Rico_Bautista_2019}. A digital album allows the users to find the different species with their relevant information and visualized using AR (see Figure \ref{fig:digitalalbum}). Such applications can unlock enthusiasm and bring people closer to nature.\par   
Nim et al. \cite{greatbarrierreef16} present a large-scale MR visualization of data about the Great Barrier Reef (GBR) and explore the impact of individual factors on the coral reef ecosystem. The visualization combines tiled displays and head-mounted displays, and dynamically presents individualized coral damage information grounded on viewers’ footprint inputs. During a tour, users provide estimates of their water and carbon footprint, based on their own typical household activities like air travel or water use. They then can see an information of GBR bleaching and crown-of-thorn sea star (COTS) outbreak, supported by immersive technologies. The authors underline that "The aim of such interactive visualization application is to increase community understandings of the complex interplay of different factors for the reef’s health"\cite{greatbarrierreef16}. \par

While the above mentioned works focus on techniques using AR, VR has also been used to promote ecological awareness. DeLeon and Berry~\cite{Deleon98virtualflorida} discuss an early (1998) attempt to create the illusion of "being there" with a projection-based 140-degree panoramic display. The installation allows spectators to navigate an airboat vehicle through a (at that time) highly realistic virtual landscape representation of the Everglades national park. Two decades later, promoting ecological awareness using VR is still a current topic as the discussion of immersive media for environmental awareness by McGinity~\cite{McGinity2018} shows.
Nelson et al. \cite{nelson2019} describe the usage of 3D HMDs for showing 360$^{\circ}$ films about coral reefs and the importance of protecting them vs. showing a unidirectional film. They examine the effects of message framing (positive vs. negative) and VR on conservation behavior and emotions. The participants' behavior was measured using donations to a conservation charity organization. The authors emphasize that VR has the potential to raise ecological awareness and to attract more people to donate.
Another current study~\cite{Markowitz2018} shows that immersive field trips into an underwater world, allowing users to experience the process and effects of rising sea water acidity, facilitates learning about climate change.
\subsubsection{Resource Efficiency}
Non-renewable energies like fossil fuels, coal and natural gas are being reduced day-by-day. They cannot be revitalized and their usage by-products and emissions are harmful to the environment. There is a clear need to reduce usage of these resources and also to promote the use of renewable sources such as bio fuels or solar and wind energy. 

Informing people about their electrical power consumption can definitely make them more vigilant as discussed in \cite{smartmeter}.
Here, authors present a proof-of-concept where sensor nodes are installed between load and power socket supply. Electrical parameters from these sensor nodes are visualized in an AR environment through the communication layer based on Internet of Things (IoT). This allows consumers to visualize in AR the consumption of the appliances through the app, become aware of the situation and make more efficient use of this resource. Mylonas et al.~\cite{MYLONAS201989} describe an IoT data AR energy visualization interface system used by students in a school laboratory. Various types of information like energy consumption, temperature, humidity and other real-time sensor data can be visualized. The authors claim that this educational activity using AR helps creating awareness among children about energy efficiency because it allows them to grasp it more easily.\par
Lu~\cite{Lu2015homeenergy} presents a system which provides eco-feedback via smart interconnected devices to remind users about their energy usage or waste. Their system synchronously visualizes the information of the physical environment with the virtual environment by using IoT enabled technologies. A 3D pet raising game rendered by a "Visualization Engine" module enables the continuous engagement of the users. The pet (an avatar) synchronizes with the activity of the user. "According to the contexts in this environment, an energy-gorging appliance will be incarnated by a monster to attack the pet. The player can turn off the energy-gorging appliance by beating the monster via the "User Interface" or by manually switching it off"\cite{Lu2015homeenergy}. Pre-processed sensor data make the system activity-aware. In a study, Lu found out that users prefer graphical to text-based visualization because the information is more comprehensive and agreeable. They also found game-based eco-feedback to be more motivating. \par
Chudikova et al. \cite{Chudikova_2019} reveal how VR and AR can help in planning, modelling and decision making in the selection of heat sources during the construction of buildings. Simulation of different heat sources can be evaluated against certain factors like price, lifespan etc. to decide upon it. Wind simulation in a 3D software or a VR walk through a complex machine room can help in the decision-making process for a suitable location for a wind turbine installation or for individual heating systems in a given space. The authors conclude that by using VR/AR technologies the heat source placement process can become clearer, easier and faster. \par
The production of renewable energy is getting more and more substantial around the world. For distant and isolated areas, where a demand for electricity is an everyday problem, photovoltaic systems and especially water pumping systems are a great value. Zenati et al.\cite{zenati2015photovoltaic} propose a distributed e-maintenance platform enriched with AR functions for a photovoltaic pumping system. A remote expert sees the scene captured by the camera in the worker's device and proposes a maintenance procedure. A corresponding marker is identified by the AR system, and the local worker gets the information as a virtual annotation that is overlayed on the marker through an HMD. The benefits that AR brings for remote assistance are further discussed in Section \ref{subsubsection:Remote assistance} of this paper.\par
All techniques supporting the construction of energy efficient products can also be considered as helping to reduce overall energy consumption. Construction tools, if designed for VR as presented by Neugebauer et al.~\cite{Neugebauer2011}, can be especially effective for this task. The authors show how a tool for understanding energy flows in a machine tool can be implemented in VR.

\subsubsection{Environmental Education}
The goal of environmental education is to increase people's understanding of nature and enabling them to attain a more ecological way of life\cite{competence14}. A person’s appreciation for nature seems relevant for motivating the search for more information about nature and environmental systems and the other way around: learning about environmental processes feeds into a person’s appreciation for nature and into involvement in environmental protection. Every individual on this planet must be educated about the different species that co-exist with us and AR/VR/MR can support that.\par
Theodorou et al.~\cite{Theodorou_2018} develop an AR based mobile application intended for teaching school students about climate change concepts and renewable energy resources. QR codes are scanned to view different learning topics. The authors conducted a test to measure knowledge gained by students before and after using the app to evaluate whether the use of the AR tool improves the learning experience. Results indicate that the increase in knowledge after using AR tool was higher compared to traditional teaching methods.\par
Huang et al.~\cite{huang2019ecologicaldata}, describe a system that translates data of an ecological model into a high-fidelity 3D model in VR. The application can serve as an educational tool and the students can achieve knowledge about climate change in a virtual forest. Two climate scenarios can be experienced on different trees, as the system retrieves the necessary information from a 3D tree database. The users can toggle different species on and off and can use a slider to easily change viewing height from ground to higher perspectives. The authors emphasize that "…creating an immersive experience of a future forest can shorten the psychological distance and increase the understanding of complex scientific data".\par 
As a part of the EcoMOBILE project Dede et al. have created a mobile-based AR game consisting of multiple modules for students to use during their field visits~\cite{ecomobile}. The Atom Tracker Module enables students to follow a carbon or oxygen atom through their environment. It can help students better understand the cycling of matter in ecosystems, such as the processes of photosynthesis and respiration. Another AR module is the Water Quality Measurement AR experience, which invites students to explore a real pond or stream. It leads students to pick up environmental probes by visiting a water measurement “toolbox”. Observations were made to investigate how well students have learned basic concepts. Results indicate better understanding while using the AR app. The main reason is increased engagement due to the enriched experience that AR provides.\par 
 
Srisuphab et al.\cite{zoo14} designed an application (ZooEduGuide) for motivating teenagers and children to learn about animals and wildlife and to raise their awareness for environmental preservation. They tested their application at a zoo. In the camera mode of the mobile phone, while using AR, points of interest (POIs) are placed on a map in real-time and the app shows the distance from the current location to the POI.\par

\subsection{Indirect Application Areas}
A large portion of the world-wide $CO_{2}$ emissions (about $20\%$) is created by the transport sector \cite{ourworldindata}. Apart from the transport of goods, these emissions are generated by people travelling either for professional (e.g. business meetings) or for personal reasons (e.g. tourism). Systems based on AR or VR that present the opportunity to do certain tasks remotely could have a positive impact in the reduction of emissions and pollution due to travel while training simulations can save resources. In this section we look deeper into some of these indirect ways in which AR/VR/MR can support the environment.

\subsubsection{Remote Collaboration}
\label{subsubsection:Remote collaboration}
Remote collaboration refers to working together, regardless of different geographic location. Remote collaboration systems have several potential impact factors on the environment. They can reduce the necessity of business travels and therefore help in a global reduction of carbon emissions. In addition, employees can be encouraged to work from home, thus reducing the total office space required, which in turn reduces the companies' environmental footprint.
Various traditional video conferencing applications have been popular among people since a few decades but they lack the personal presence of a physical meeting. Advanced AR/VR/MR systems which create shared spaces that enable person embodiment and haptic interaction could become a more convincing alternative to travelling to distant conferences and meetings and help in reducing travel costs, time, carbon emissions and overall office space required. \par

We identify three main components in remote collaboration systems based on AR/VR/MR: \textit{Environment}, \textit{Avatars} and \textit{Interaction}. \par
\textit{Environment.} The virtual environment stimulates the sensory impressions of its users such as vision, hearing, touch or smell. While older systems experimented with the stimulation of many sensory impressions, more recent systems tend to rely on audiovisual stimuli only. As an example, Morton Heilig created an environment which featured 3D visuals, stereo sound, seat vibration, wind from fans and olfactory cues using aromas in the early sixties \cite{heilig1962sensorama}. Marker based AR systems tend to have haptic feedback with tangible interfaces such as turntables \cite{shen2006framework, shen2008collaborative} or a pen \cite{shen2008product}. However, marker based systems are becoming obsolete due to the increasing maturity of AR tracking technology and therefore are developing a trend to use only audiovisual stimuli. The authors of \cite{schafer2019towards} use multiple static panorama images to create a virtual meeting room where multiple people can participate simultaneously. It features simple presenting tools such as a virtual projector or TV. A live panorama video stream is used by Lee et al. \cite{lee2017mixed} to create a wearable MR remote collaboration system. 

\textit{Avatar.} An avatar represents the user in virtual space (without necessarily resembling the appearance of the user) and is key to enabling communication to other users in a collaborative virtual environment. Monahan et al. \cite{monahan2008virtual} implemented a remote collaboration system for knowledge transfer, which allowed online lectures and students having online group meetings. The system featured avatars which could use gestures (e.g. raise hands) for communication besides audio and text chat. A system with an adaptive avatar was created by Piumsomboon et al. which featured a miniature avatar which was able to transmit nonverbal communication cues \cite{piumsomboon2018mini}.

\textit{Interaction.} The possibility to interact with the environment and other users is another important component of remote collaboration systems. Systems developed for product design such as \cite{shen2010augmented, shen2008product} usually implement shared 3D object manipulation, including viewing, changing and annotation of 3D objects. Remote collaboration systems are also popular in architectural design, where systems usually integrate the viewing of houses, rooms and furniture while allowing the users to discuss either via audio or integrated tools such as drawing in the virtual environment \cite{hsudesign, chowdhury2019laypeople, hong2019architectural}. Many professional remote collaboration systems such as \cite{VirBELA, MozillaHubs, EngageVR, MeetInVR, Glue, breakroom} typically include screen sharing and media sharing as interaction possibilities.
\par
Due to the COVID-19 outbreak, many events and conferences were held completely virtual. Some events utilized professional VR remote collaboration systems such as the IEEE VR 2020 conference. The organizing committee used a modified version of Mozilla Hubs \cite{MozillaHubs} which is a professional social VR platform. During this event, the platform featured virtual rooms with up to 20 persons simultaneously in VR. Furthermore it enabled 3D placement of custom content such as posters and presentations. The poster session of the event was held completely in VR. \par
\begin{figure}[h]
\centering
\includegraphics[width=0.9\columnwidth]{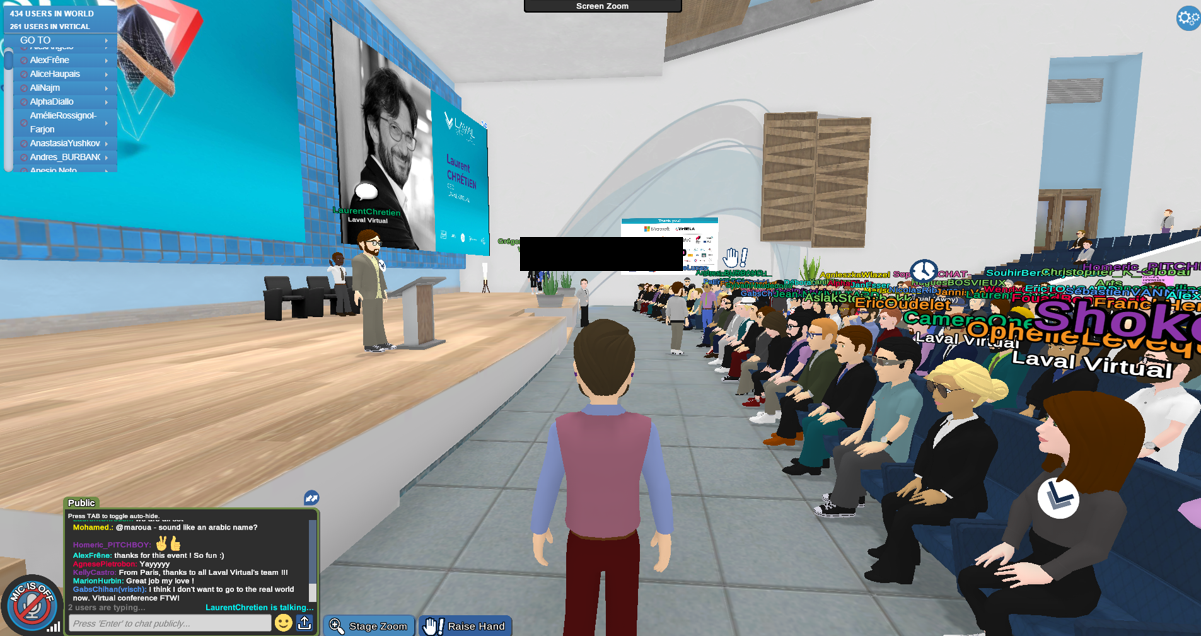}
\caption{The Laval Virtual 2020 held completely in VR. One of the authors took this screenshot while attending the event with the avatar that can be seen in the center of the image (nickname anonymized).}
\label{fig:lavalvirtual}
\end{figure}

Another example is the Laval Virtual, an event usually attended by around 10,000 participants and 150 speakers with content covering three days. Due to the special circumstances during the COVID-19 outbreak, the 2020 event was held in VR (See Figure \ref{fig:lavalvirtual}), presumably preventing many business trips and thus contributing to $CO_2$ savings. The professional platform used here was VirBELA \cite{VirBELA}, a platform specialized for large scale events in VR. The presenters shared their screen virtually to all attendees and held the presentation with an avatar in a 3D environment. Each person in the VR event was visible as an avatar and the presenter got in front of a podium just as it would be in the physical event. Each attendee had his own avatar and was part of the virtual conference. \par
Additionally to the meeting and design use cases there are remote collaboration systems which are used for remote assistance, which we address in the next section.

\subsubsection{Remote Assistance}
\label{subsubsection:Remote assistance}
Remote assistance or remote support is one of the most heavily explored AR use-cases, especially in industrial scenarios \cite{Gallala_2019}. The core idea of such applications is that a live connection can be established between an on-site user and a remote expert-maintainer. Using AR annotations and optionally audio or text, the expert can guide the user into performing a challenging task. Such processes have the potential of saving time and most importantly allow specialist maintainers of devices to avoid long-distance travel to perform on-site repairs. Apart from industrial maintenance use-cases the idea is also of interest for general customer support for home devices.\par
Several approaches addressing the topic have been introduced, from early research to commercial systems. Early approaches relied on augmentations that are registered in 2D \cite{masoni2017supporting}, but it soon became apparent that 3D registration of the target is needed for such systems. 3D object tracking of the maintenance target using an object model as reference is a viable option presented in \cite{DoFarticle}. Alternatively, the geometry of the environment needs to be reconstructed in real-time by a SLAM-based system in order to place annotations in 3D space \cite{zillner2018augmented}.\par
Remote support is one of the most commercially advanced AR use-cases. Nowadays, several companies offer such services, either for see-through AR on head mounted displays (HMDs) or video see-through on mobile devices \cite{scopear,vuforiachalk,reflekt}. Despite the large investment, it is for now not clear to which extent such systems are already used in practice in industrial maintenance.\par
Currently, the more challenging task seems to be the capturing of the knowledge of the remote expert and its embedding in an intelligent system that is able to provide instructions for its maintenance or that of other systems. In the work of Petersen et al. \cite{petersen2013real}, the focus is placed on automating knowledge capture from workflow videos, while Rambach et al. \cite{rambach2017things} present the idea of IoT objects that carry their own AR content that can be shared presented upon connecting to the objects. Furthermore, hand tracking for modelling the interaction of users with objects will be able to add value to maintenance applications \cite{hampalihonnotate}, as well as recognizing objects and estimating their pose at different states \cite{su2019deep}. 

\subsubsection{Remote Tourism}
Tourism beyond attraction destinations and entertainment also contributes to economic growth of a country by generating revenues, creating more employment opportunities and supporting the development of infrastructure. As an example, the tourism industry in Europe had around 1.5 billion nights spent at tourist accommodation establishments and over 200 billion euros spent by tourists in 2018 according to Eurostat \cite{eurostat}. The relationship of tourism and the environment is complicated. As mentioned above it has positive effects, but it can also have a negative influence on the environment. It can be the cause of depletion of water and land resources and a source of pollution. Thus, tourism can gradually damage the environment which it depends on \cite{U.2003}. VR and AR technologies provide alternative forms of visiting famous destinations by bringing out the sense of presence without impacting the natural environment while also saving time, energy and costs. \par  
In the work of Fritz et al. \cite{Fritz2005EnhancingCT}, authors illustrate a 3D visualization device developed using AR technologies and the tourist binoculars concept. With the help of this device users can view a remote place and obtain personalized multimedia information about it. Such devices can be used to get attractive and educative information about historical buildings, museums, art galleries and national parks. Furthermore, in national parks visitors can not only gain facts and information on the current site but also view fossil remains and traces of living beings in distant and unreachable locations. Another recent example is the work of \cite{jung2020meet} that offers a remote cultural heritage experience of the Jeju island in Korea through AR and VR.
According to Moiseeva et al. \cite{Moiseeva20195313}, remote tourism using AR provides opportunities to disabled people to visit heritage sites, castles, museums and other inaccessible locations.

\begin{figure}[ht]
\centering
\includegraphics[width=0.9\columnwidth]{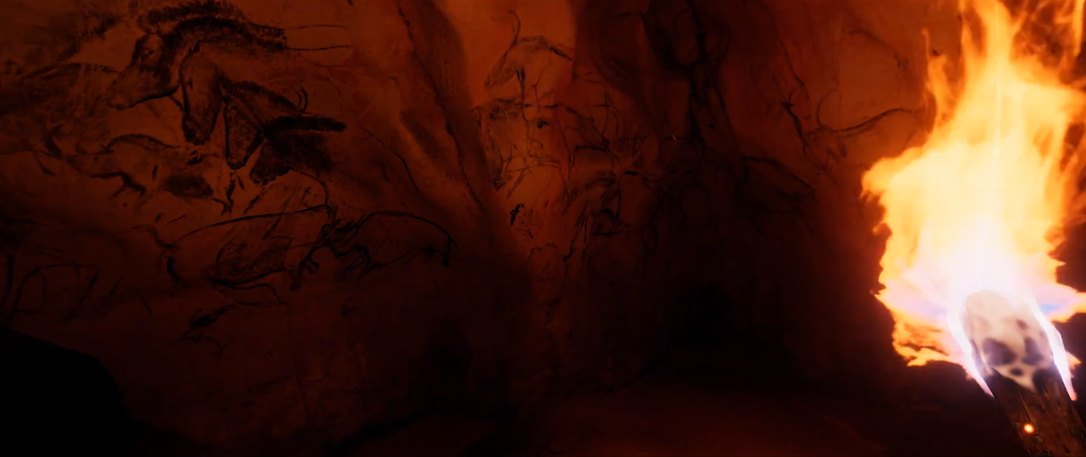}
\caption{\label{fig:chauvetcave}VR scenario of Chauvet Cave in Ardèche, France. Instead of visiting the place physically, it can be explored fully in VR \cite{dawnofart}. Screenshot taken by one of the authors while using the VR application.}
\end{figure}

As an example, the Google Arts \& Culture project \cite{googleartsandculture} enables people from all over the world to experience an immersive world in AR and VR based on real locations. Many museums and galleries are already part of such projects and make their exhibited works available for preservation purposes. To mention one, the Chauvet Cave in Ardèche, France, is home to some of the oldest cave drawings in the world according to \cite{chauvet1996chauvet}. There are AR and VR applications to experience the cave in a completely virtual way. In AR, objects can be placed in the augmented world, such as pictures or parts of the cave with interesting wall paintings. A fully immersive experience in VR offers a full 3D reconstruction of the cave with interactive elements such as a torch to illuminate the cave and points of interest (See Figure \ref{fig:chauvetcave}). Freely available applications such as these, not only improve the preservation of cultural heritage but could also be an effective environmental measure since travel is replaced by virtual tours.\par

Although remote tourism can be a substitute for actual traveling, the work of Bogicevic et al. \cite{bogicevic2019virtual} suggests tourism brands to create VR platforms as marketing strategies . This turns VR in tourism into a double-edged sword. On the one hand, travel is prevented, but on the other hand the desire to travel can be awakened~\cite{Drengner2019}. During the current travel-ban that most people are experiencing due to the COVID-19 pandemic, remote tourism can also be an opportunity for creating some revenue for touristic attractions that are not accessible.

\subsubsection{Simulation/Training}
\label{sec:training}
Similarly to remote tourism, techniques that simulate certain environments and situations can also allow saving resources and reducing travel and emissions. Examples thereof are VR simulations of driving automobiles or aircrafts that can be used for training new drivers/pilots as in the work of Riegler et al. \cite{riegler2019autowsd} and Bruguera et al.  \cite{bruguera2019use}. Simulations for other forms of training such as complex machinery or even training of personnel for critical situations is shown in the work of Gavish et al. \cite{gavish2015evaluating} and Engelbrecht et al. \cite{engelbrecht2019swot}.

The level of success of such applications depends highly on the quality of the simulation. While applications such as remote tourism mostly provide visualization with limited interaction, other aforementioned simulation applications require much more detailed interaction and feedback mechanisms that rely on fully understanding body motion and interaction as well as haptic feedback. \par

Another example is VR for high risk training in particularly dangerous environments. For example, firefighters train in a nearly controlled environment, but those training conditions are still very dangerous since they must prepare for dangerous real life situations. This training is not only life threating but also facilitates the destruction of the environment. According to Evarts et al. \cite{evarts2018united} and Fahy et al. \cite{fahy2018firefighter}, there are over 8380 injuries and ten deaths from training accidents in the United States in 2017. Unfortunately, these dangerous training scenarios are still necessary today to train and educate firefighters. These numbers promote virtual training scenarios since they do not include preparing and moving in dangerous environments. There are companies that specialize in virtual training scenarios, e.g. Flaim Systems \cite{flaimsystems} offers many realistic VR training scenarios for firefighters. In order to provide even more immersion, special hardware is also used for such training scenarios. For example a combination of real equipment like a firefighter suit, custom made hardware like a water hose trackable by VR and a heat west. Often a portable computer is stowed in a backpack and is connected to the VR HMD to provide the necessary mobility. The work of Engelbrecht et al. \cite{engelbrecht2019swot} gives a summary of VR training scenarios for firefighters. \par

Virtual training and simulation scenarios can also be used as marketing strategies whereby the environment benefits from these activities as well. In the home improvement sector, many people are not confident enough to do advanced or even simple tasks. For this, Lowe (a large home improvement store in the USA) created \textit{Holoroom How To} \cite{holoroomhowto} and offers VR training in stores. The VR training includes mixing paint/cement, painting walls, tiling bathrooms and other use cases. Lowe uses this training to encourage customers to try for themselves and therefore generate more revenue. Another example is \textit{Holoroom Test Drive} \cite{holoroomtestdrive} which is used as a try-before-you-buy model to enable customers to try out dangerous products like chainsaws or hedge trimmers in a safe, virtual space. It uses custom made hardware for a more educational and immersive experience. Simulations such as these help reducing the number of returned goods and thus also the ecological footprint.\par
The same applies to AR shopping apps. As an example, the IKEA Place app launched in September 2017 and has over 3.200 products in its portfolio according to Alves et al. \cite{alves2020intention}. One of the main advantages for customers is that they can see the real dimensions of products they want to buy and thus avoid unnecessary returns if the product does not fit as expected. Studies from Alves et al.\cite{alves2020intention} and Dacko et al. \cite{dacko2017enabling} have shown that users of AR placement apps feel greater confidence and greater purchasing convenience with an AR furniture placement app,  which ultimately leads to less returned goods as well. The mentioned applications are also reducing the emissions caused by transportation due to trips not made if the products of interest have not convinced virtually. \par

\section{Discussion}
\label{section:Ideas}
Examining the outcome of our search, we confirm our initial notion that ecological applications of AR/VR/MR are not yet widely researched or applied in products. Although this survey covers a considerable part of the existing literature, it was only possible to give an overview of the topics by focusing on important works and explaining their relevance to nature and the environment. For example, remote collaboration generally has a positive impact on the environment, regardless of the actual implementation details. To the best of our knowledge, there is no remote collaboration system which was developed for the sake of a positive impact on the environment. \par
In the category \emph{direct} we identified papers that explicitly discussed applications for environmental education, monitoring, awareness and resource efficiency. Statistical analysis of these publications (Section \ref{section:Stats}) reveals some interesting insights even though the number of existing papers is limited to only 28. In the classification by topic (Figure \ref{fig:stats}), environmental awareness appears to be significantly more popular than other categories. It is surprising that only 3 publications were found in the category environmental education when considering that education in general is arguably a heavily explored field of AR/VR/MR applications. The classification of publications by year shows an almost complete absence of related publications before 2011 as well as a possible developing trend in the last years (2018 and after). Finally, the fact that most publications were not published in dedicated AR/VR/MR conferences and journals shows that the topic has possibly not yet received significant attention by the AR/VR/MR community. \par 
Looking more in detail at the publications we can draw further insights. Some of these works try to actively involve many users, as active participation is very important for learning, understanding, and engagement. This results in applications using common hardware such as smartphones and tablets. Therefore, many applications focus on technologies running on those devices. The usage of HMDs can be found more in applications for experts and in the industry field, which is also the main target group for the latest HMDs (e.g. Microsoft Hololens \cite{hololens}). Nevertheless, it should be noted that the use of AR only for visualization or with limited interaction properties is a common pattern in this category. However, user studies performed show increased immersion or educational results in these cases. Many approaches still rely on tracking using markers as targets, a concept which is not fully scalable to all application scenarios.
Efficient tracking and unknown scene understanding at a geometric as well as semantic level is still the main challenge that has to be overcome in order to facilitate AR applications in all covered areas. Although important steps in this direction have already been made in recent years (see for example Microsoft Hololens \cite{hololens}), significant further advancement is expected in the future. Understanding dynamic scenes, human motion and hand gestures is additionally a current topic of importance.\par 
It is also important to mention that included publications mostly refer to augmenting the senses of sight and hearing while the other three - the sense of touch, taste and smell, are not considered. For example, the smell of fire could be very interesting for forest fire simulations as part of a climate change and for achieving better immersion. Although there is a small niche for some commercial products, most hardware for such applications is in a development phase, and the usage of our five senses in AR/VR/MR applications could be seen as a direction for the far future.  \par 
Some ideas for use cases in everyday life are discussed by McGinity \cite{McGinity2018}. He describes the usage of AR in a supermarket scenario for visually informing people about how far a product has travelled or how the choice of recycled products like paper towels could help protecting the forests. For these cases he suggests to visualize a forest regrow in AR. This type of AR usage is possible with the current state of the art and could be developed in the near future, perhaps to complement some other applications for products. \par 
In the category \emph{indirect} we covered the topics remote collaboration, remote assistance, remote tourism, and training with their respective applications. These applications have a positive impact on nature and the environment, but were developed and evaluated for other reasons. During the survey, we found that these applications have achieved different readiness levels. For example, basic remote support applications are less challenging and have been already integrated in commercial products, while remote collaboration systems based on AR/VR/MR have larger challenges that they need to overcome and are thus still at a research level. Especially during times when physical distancing is advised, the research in remote collaboration systems with AR, VR or MR technology will focus more on transferring nonverbal communication between users to substitute physical meetings.\par
During the COVID-19 outbreak which started in December 2019, a large part of the worldwide population has been required to remain at home as a protective measure. All remote collaboration systems and remote tourism scenarios thus have become even more important for the support of economies. This should be seen as a great opportunity to show how much can be achieved remotely and to reduce travel and emissions in the future and thus supporting the environment. Of course, not all human contact can or should be replaced by these technologies. They can however provide viable alternatives to some travels.  \par
Other possible environmental application areas are training scenarios, which help to reduce the environmental destruction. An example described in section \ref{sec:training} is the training of firefighters, where environmentally harmful resources are used by starting real fires and extinguishing these fires with chemicals. Virtual training scenarios are already quite advanced and often use custom hardware to support realism and immersion, making them a good substitute for real world training. \par
AR applications in the e-commerce sector can help the environment, reducing the ecological footprint of companies by minimizing return of goods. There are already many applications which let the user place objects (e.g. furniture in one's own home) and check if the color fits or measure the size of those objects to avoid mispurchases.
\section{Conclusion}
\label{section:Conclusion}
This survey explores AR/VR/MR applications and concepts that have a relation to nature, environment preservation and resource efficiency. Some work has been developed for direct impacts on nature. Other areas have an indirect impact on nature, such as remote collaboration systems, which are developed for the purpose of remote working or socializing together. These systems were primarily developed for knowledge transfer or virtual meetings, but they have an indirect impact on nature by reducing travel and office space. Therefore, the survey results are classified into applications that have a direct relation to ecology, and applications that can provide an indirect support on ecological goals. To the best of our knowledge there is no comparable survey with similar goals. Results of the survey confirmed that there is only a limited amount of work in AR/VR/MR with a direct ecological relation. After a further categorization, the benefits of AR/VR/MR across a wide variety of domains like education, energy consumption, ecological sensitization and remote collaboration are discussed. In retrospective, the main purpose of this survey was to explore an AR/VR/MR application topic that is not often considered, increase awareness in the scientific community, and create a direct reference to the existing work for the future. Considering the indirect cases in the classification, the search area has expanded considerably. This serves as an indicator that there is a real potential for environmental protection in the use of AR/VR/MR. For some applications this potential can be realized in the near future, while for others a significant advancement of enabling technologies is still required.

\vspace{6pt} 



\authorcontributions{Conceptualization, J.R.; methodology, J.R., investigation, J.R.,G.L.,A.S. and R.B.; formal analysis, J.R.,G.L.,A.S.; writing, J.R.,G.L.,A.S.,A.W. and R.B.; review and editing, J.R.,A.W.,D.S.; funding acquisition, A.W.,D.S.; All authors have read and agreed to the published version of the manuscript.  }

\funding{Parts of this work have been performed in the context of project SAARTE (Spatially-Aware Augmented Reality in Teaching and Education). SAARTE is supported by the European Union (EU) in the ERDF program P1-SZ2-7 and by the German federal state Rhineland-Palatinate (Antr.-Nr. 84002945). This work was also supported by the Bundesministerium für Bildung und Forschung (BMBF) in the context of ODPfalz under Grant 03IHS075B}


\conflictsofinterest{The authors declare no conflict of interest. The funders had no role in the design of the study; in the collection, analyses, or interpretation of data; in the writing of the manuscript, or in the decision to publish the results.} 


\reftitle{References}


\externalbibliography{yes}
\bibliography{references}




\end{document}